\documentclass[superscriptaddress,prb,preprint,amsmath,amssymb]{revtex4}

\usepackage{graphicx,natbib}

\usepackage{calc,graphicx,float,amsmath}

\usepackage[paper=letterpaper,
            includefoot,
            marginparsep=.05in,       
            margin=0.75in,               
            includemp]{geometry} 

\usepackage{bm}  

\def\sx{ {s^{\rm ex}} }

\def\DBD{ {1-D_{\text{BD}}/D_0} }

\def\kB{k_{\text{B}}}

\def\DRosen{{D_{\text{MD}}\rho^{1/3}(m/\kB T)^{1/2}}}

\usepackage{varioref}
\begin{document}





\title{Mapping between long-time molecular and Brownian dynamics}

\author{Mark J. Pond} 
\affiliation{Department of Chemical Engineering,
  The University of Texas at Austin, Austin, TX 78712.}

\author{Jeffrey R. Errington} 
\affiliation{Department of Chemical and Biological Engineering,
  University at Buffalo, The State University of New York, Buffalo,
  New York 14260-4200, USA}

\author{Thomas M. Truskett} \email{truskett@che.utexas.edu}
\thanks{Corresponding Author} {}
\affiliation{Department of Chemical Engineering, The University of Texas
  at Austin, Austin, TX 78712.}
%

%
%
%
  
  
\begin{abstract}
  We use computer simulations to test a simple idea for  mapping between long-time self diffusivities obtained from molecular and Brownian  dynamics.  The strategy we explore is motivated by the  behavior of fluids comprising particles that interact via inverse-power-law pair potentials, which serve as good reference models for dense atomic or colloidal materials. Based on our simulation data, we present an  empirical expression that semi-quantitatively describes the ``atomic'' to ``colloidal'' diffusivity mapping for inverse-power-law fluids, but also  for model complex fluids with  considerably softer (star-polymer, Gaussian-core, or Hertzian) interactions. As we show, the anomalous structural and dynamic properties of these latter ultrasoft systems pose problems for other strategies designed to relate Newtonian and Brownian dynamics of hard-sphere-like particles.
\end{abstract}
\maketitle

Computer simulations and statistical mechanical theories have long served as invaluable tools for understanding relaxation processes that occur in fluid systems that range from  molecular liquids to complex suspensions of Brownian particles.~\cite{Likos2001effective-interactions} Depending on the resolution of the interparticle interactions used to model these systems, some descriptions of the microscopic dynamics are more appropriate to adopt than others. Simplistically, Newtonian (i.e., classical molecular) dynamics (MD) is suitable when the particles of interest are described by force fields with molecular-scale resolution. Modeling the motion of suspended Brownian particles, on the other hand, typically calls for coarser interparticle forces and dynamics that approximately account for the effects of the solvent. \cite{Lowen1994Melting-freezin,Ermak1975A-Computer-Sim,Brady1985The-rheology, Hoogerbrugge1992Simulating-Micr, Malevanets1999Mesoscopic-mode}

Although the short-time behavior of a given model depends sensitively on its microscopic dynamics,\cite{Heyes1994Molecular-and-Brownian} there is evidence that under certain conditions--e.g.,  dense fluids near the glass transition--the qualitative long-time behavior is largely independent of those details.\cite{Lowen1991Brownian-dynami, Gleim1998How-Does-the-Re} With this in mind, it is perhaps not surprising that  researchers often  select the type of microscopic dynamics to use in  simulations  based on other considerations, such as computational efficiency. A common example is adopting  classical MD to explore the long-time dynamic behavior of  model complex fluids with coarse, effective interactions, ignoring the kinetic role of the implicit ``fast'' degrees of freedom, e.g, solvent.\cite{Stillinger1978Study-of-meltin, Foffi2002Evidence-for-an,Puertas2003Simulation-stud, Mausbach2006Static-and-dyna,Moreno2007-Cluster, Krekelberg2009Gaussian-dynamics, Pamies2009Phase-diagram, Pond2009-Composition-and-conc,Berthier2010arXiv-Increasing}
Although the  qualitative trends provided by these MD simulations have been  valuable, there is still the question of how to map between the long-time MD data of such simulations and that which would have  been produced assuming a different  type of microscopic dynamics. 

Here, we take a first  step toward addressing this issue.  Specifically, we present computer simulation results that test a  simple heuristic approach for mapping between long-time self-diffusion coefficients obtained from Newtonian and Brownian (i.e., overdamped Langevin) dynamics (BD). The method  follows from a seemingly naive hypothesis that ``what matters" in such a mapping can be deduced from the behavior of a fluid of particles that interact via an inverse-power-law (IPL) pair potential~[$\mathcal{V}(x)=\epsilon x^{-\mu}$], where $x=r/\sigma$, $r$ is the interparticle separation, $\epsilon$ and $\sigma$ represent characteristic energy and length scales of the interaction, and $\mu$ determines the steepness of the IPL repulsion. The IPL model is a natural reference system for  dense atomic or colloidal fluids, whose static and dynamic properties are dominated by the repulsive part of their interactions. Furthermore,  one can show\cite{Rosenfeld1977Relation-betwee, Hoover1991Computational-S, Gnan2010Pressure-en} that a one-to-one relation must exist between the following dimensionless representations of the long-time diffusivities of an IPL fluid associated with the two types of dynamics: $D_{\rm MD} \rho^{1/3}\sqrt{m/k_{\rm B}T}$ and $D_{\rm BD}/D_0$.  Here, $\rho$ is the number density, $m$ is the particle mass, $k_{\rm B}$ is the Boltzmann constant, and $T$ is temperature. $D_{\rm MD}$ ($D_{\rm BD}$) represents the long-time self diffusivity obtained  from MD (BD) trajectories, respectively, and $D_0$ is the value of $D_{\rm BD}$ in the  dilute ($\rho \rightarrow 0$) limit. Based on previous work,\cite{Rosenfeld1999A-quasi-univers,Lange2009Comparison-of-s,Pond2011Generalizing-Rosenfelds} it is clear that the relationship between $D_{\rm MD} \rho^{1/3}\sqrt{m/k_{\rm B}T}$ and $D_{\rm BD}/D_0$ for IPL fluids is approximately independent of $\mu$ for $\mu>4$. Hence, a more precise statement of our aforementioned hypothesis is that the quasi-universal IPL\ mapping relationship can also be used  to estimate  $D_{\rm MD}$ from $D_{\rm BD}$ (or vice versa) for other types of fluids, perhaps including those with very different types of interactions and physical properties.  

To test this hypothesis, we carry out MD and BD\ simulations that probe  the long-time dynamics  of fluids of particles that interact via IPL potentials with $\mu=8$, $10$, $12$, $18$, and $36$. We also explore the behavior of fluids with particles interacting via ultrasoft Gaussian-core\cite{Stillinger1976Phase-transitio}~[$\mathcal{V}(x)=\epsilon \exp\{-x^2\}]$, Hertzian\cite{Pamies2009Phase-diagram}~[$\mathcal{V}(x)=\epsilon (1-x)^{5/2}$], and effective star-polymer\cite{Likos1998Star-Polymers} [$\mathcal{V}(x)=A\{-\ln x+B\}$  for $x<1$, and $\mathcal{V}(x)=ABx^{-1}\exp\{(B^{-1}-1)(1-x)\}$ for $x\ge1$] potentials. Here, $B=f^{1/2}/2$, $A=(20/9) k_{\rm B}T(B^{-1}-1)^3$, and $f$ is the star polymer arm number. The latter three model systems have received considerable attention recently in the theoretical soft matter literature,\cite{Likos2001effective-interactions,Foffi2003Structural-Arre,Pamies2009Phase-diagram} and represent stringent  test cases for our approach (and others) due to their distinctive dynamical trends; e.g., each exhibits a wide range of conditions where self diffusivity anomalously increases with \textit{increasing} particle density, as opposed to the behavior of  IPL fluids.
  The MD and BD simulations that  we present here  cover much of the computationally accessible phase space of these model fluids, including both equilibrium and moderately supercooled conditions ($\sim10^3$ state points in total).
As we show below, a potentially useful outcome of our analysis  of this extensive data set is an empirical analytical equation that  semi-quantitatively relates $D_{\rm MD}$ and $D_{\rm BD}$ for these systems.  

The  MD  simulations of our study generate dynamic trajectories by solving Newton's equation of motion using   the velocity-Verlet algorithm in the microcanonical ensemble.\cite{Allen1987Computer-Simula} For IPL, Hertzian, and star-polymer fluids, they contain $N$ particles ($N = 1000$, $4000$, and $4000,$ respectively) and use integration time steps of $0.00005 \sqrt{m \sigma^2 / \epsilon}$ , $0.01 \sqrt{m \sigma^2 / \epsilon,}$ and 0.001 $\sqrt{m \sigma^2 / k_{\rm B}T}$, respectively.  The BD simulations presented here generate trajectories by solving the  Langevin equation in the high-friction limit using the conventional Brownian (Ermak) algorithm.\cite{Allen1987Computer-Simula, Ermak1975A-Computer-Sim} For the star-polymer fluid, they contain  $N=4000$ particles, and use a time step of $\Delta t = 0.1 \tau_B,$ where $\tau_B=mD_0/k_{\rm B} T$ and $D_0=0.001\sqrt{\sigma^2k_{\rm B}T/m}$.  All simulations use a periodically replicated cubic cell with reduced volume $V/\sigma^3$ determined by the  number density $\rho \sigma^3=N \sigma^3/V$ of interest. For the IPL, Hertzian, and star-polymer systems, the pair potentials are truncated at $r_{cut} = 1.4 - 3.9 \sigma$ (depending on $\mu$)\cite{Pond2011Generalizing-Rosenfelds}, $1\sigma,$ and $3.6 \sigma$ respectively. Long-time tracer diffusivities from both MD\ and BD simulation trajectories are computed from the average mean-squared particle displacements via the Einstein relation. For the analysis presented below, we also include some previously reported MD simulation data\cite{Krekelberg2009Gaussian-dynamics} for  the Gaussian-core fluid and BD simulation data\cite{Pond2011Generalizing-Rosenfelds} for the IPL, Gaussian-core, and Hertzian fluids. The methods used in those studies are the same as those described above, and the required simulation parameters can be found in the original papers.

\begin{figure}[t]
\centering
\includegraphics[width=2.8in]{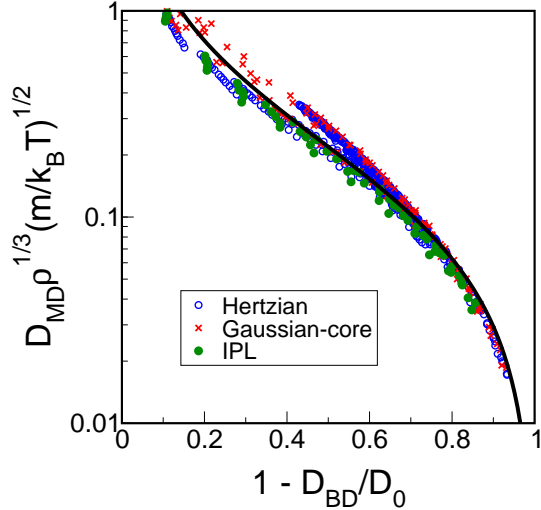}
\caption{Reduced long-time diffusivity from MD simulations $\DRosen$ versus the fractional reduction in the long-time diffusivity from BD simulations (relative to the dilute value), $\DBD$.  The blue circles correspond to the Hertzian fluid at temperatures ranging from $0.002-0.02~\epsilon/\kB$  and densities that range from $0.2 - 8.0~\sigma^{-3}$. The red circles correspond to the Gaussian-core fluid at temperatures ranging from $0.004- 0.2~\epsilon /\kB$ and densities that range from $0.01 - 1.0~\sigma^{-3}$. The green circles correspond to IPL fluids with exponents, $\mu$, ranging from 8 to 36 and values of the parameter  $\rho \sigma^3 (\epsilon/k_{\rm B} T)^{3/\mu}$ that span the respective equilibrium fluid phases. The curve is a least-squares fit to the data for the three systems, and is given by eq.~\ref{eq:BDMDFit} with   $c_1=3.3176$ and $c_2=2.6645$.}
\label{fig:TotalCollapse}
\end{figure}

Before examining the simulation data of the various model fluids discussed above, we first consider what should generally be expected about the relationship between the dimensionless diffusivities, $D_{\rm BD}/D_0$ and $D_{\rm RMD} \ \equiv D_{\rm MD} \rho^{1/3}\sqrt{m/k_{\rm B}T}$. For example, to leading order in $\rho$, we know  that $1-D_{\rm BD}/D_0 \propto \rho$ and $D_{\rm MD}^{-1} \propto \rho$, which together imply that  $1-D_{\rm BD}/D_0 \propto D_{\rm RMD}^{3/2}$ in this limit.   Furthermore, there is also evidence\cite{Lowen1991Brownian-dynami, Gleim1998How-Does-the-Re} suggesting that $D_{\text{BD}}\propto D_{\text{MD}}$ for supercooled liquids near the glass transition ($D_{\text{MD}}$, $D_{\rm BD}\rightarrow 0$). Since $D_{\text{MD}}$ necessarily shows pronounced variations with small changes in $\rho$ or $T$ under these latter conditions, we also have $D_{\rm BD}/D_0 \propto D_{\rm RMD}$. The following expression,\begin{equation}
\DBD =(1+c_{1} D_{\rm RMD} + c_{2} D_{\rm RMD}^{3/2})^{-1}
\label{eq:BDMDFit}
,\end{equation}is an example of a\ simple heuristic functional form that interpolates between the  aforementioned characteristic ``fast" and ``slow'' limiting behaviors. Below, we examine how well eq.~\ref{eq:BDMDFit} can describe the simulation data for a variety of fluids comprising hard to ultrasoft particles if $c_1$ and $c_2$ are treated as constants.  

  Computer simulation data of $D_{\rm RMD}$ plotted versus $1-D_{\rm BD}/D_0$ for the IPL, Gaussian-core, and Hertzian fluids are presented in  Fig.~\ref{fig:TotalCollapse}. The data span the $k_{\rm B} T/\epsilon-\rho \sigma^3$ plane (details in the caption), characterizing the relationship between MD and BD long-time diffusivities for these fluids in    their equilibrium and moderately supercooled states.  Also presented in Fig.~\ref{eq:BDMDFit} is a least-squares fit of the data using the form provided by eq.~\ref{eq:BDMDFit}. As can be seen--despite the distinctive non-monotonic dynamic trends of the Gaussian-core and Hertzian fluids as a function of density\cite{Mausbach2006Static-and-dyna,Wensink2008Long-time-self-,Krekelberg2009Gaussian-dynamics,Wensink2008Long-time-self-,Pamies2009Phase-diagram,Krekelberg2009Generalized-Rosenfeld,Pond2011Generalizing-Rosenfelds} --the data qualitatively behave as the equation predicts.  In fact, as is illustrated in Fig.~S1$\dag$,
more than 98\% of the simulation data for $D_{\rm BD}$ for each of these model fluids are within 20\% of the eq.~\ref{eq:BDMDFit} estimation based on the simulated $D_{\rm MD}$. Furthermore, to the extent that  the data shown in Fig.~\ref{eq:BDMDFit} reflect quasi-universal behavior, it suggests that the well-known, empirical dynamic freezing criterion for colloidal fluids ($D_{\rm BD}/D_0 \approx 0.1$),\cite{Lowen1993Dynamical-crite} has an analog in atomic systems ($D_{\rm RMD} \approx .03$).  

\begin{figure}
\begin{center}
\includegraphics[width=3in]{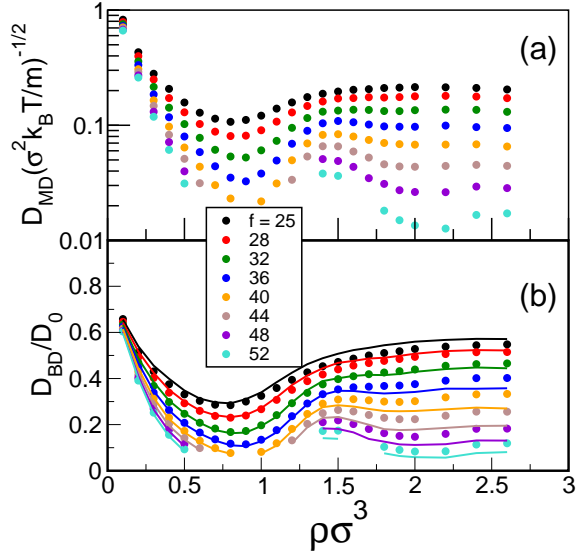}
\caption{Reduced long-time self diffusivity [$D_{\rm MD} (\sigma^2 k_{\rm B} T/m)^{-1/2}$ and $D_{\rm BD}/D_0$] for the star-polymer fluid plotted versus reduced density $\rho \sigma^3$ from MD and BD simulations, respectively Fluids of stars with different arm numbers  ranging from $f=25-52$ (top to bottom) and reduced densities $\rho \sigma^3=0-2.6$ are shown. (a) Results from MD simulations. (b) Results from BD simulations (symbols) and estimates (curves) based on simply substituting the $D_{\rm MD}$ data of panel (a) into eq. \ref{eq:BDMDFit} with constants given in the caption of Fig~\ref{fig:TotalCollapse}.}
\label{fig:StarPred}
\end{center}
\end{figure}

\begin{figure*}[t]
\begin{center}
\includegraphics[width=5in]{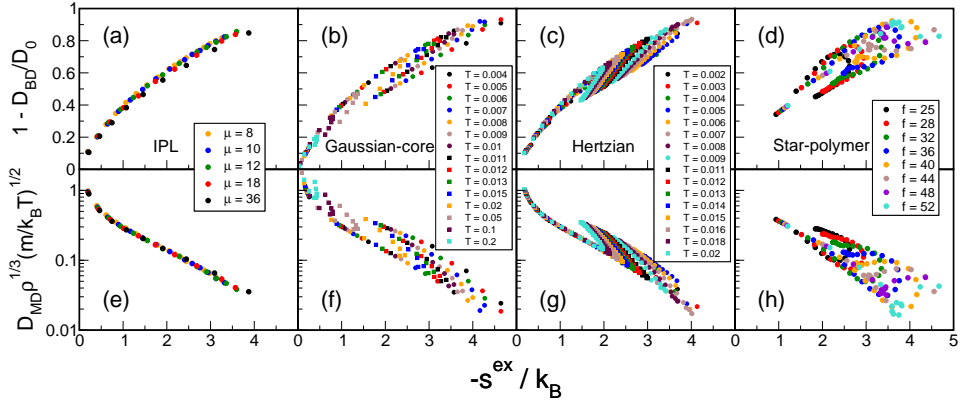}
\caption{Fractional reduction in the long-time diffusivity from BD simulations (relative to the dilute value), $\DBD$ [panels (a)-(d)], and reduced long-time diffusivity from MD simulations $\DRosen$ [panels (e)-(h)], plotted versus negative excess entropy per particle, $-\sx/k_{\rm B}$. Dynamic data corresponds to that  of Fig.~\ref{fig:TotalCollapse} and~\ref{fig:StarPred}, and $-\sx$ data were computed via the free-energy-based simulation techniques discussed in the text. }
\label{fig:sxD}
\end{center}
\end{figure*}

As an illustration of how eq.~\ref{eq:BDMDFit} might  be further used, we consider the star-polymer fluid\cite{Likos1998Star-Polymers} mentioned above. The soft, logarithmic repulsive interactions of this model are known to produce highly non-monotonic dynamic trends; e.g., diffusivity show two minima as a function of $\rho \sigma^3$ (see Fig.~\ref{fig:StarPred} of this paper and Fig.~1 of Foffi et al.~\cite{Foffi2003Structural-Arre}). When a model fluid displays such nontrivial behavior with one type of microscopic dynamics (e.g., MD), it is not obvious a priori whether the trends  will necessarily be reflected when another type (e.g., BD) is employed. In fact,    an earlier investigation of this system\cite{Foffi2003Structural-Arre} made a point to report dynamic results from  both types of simulations. What Fig.~\ref{fig:StarPred}b illustrates that is that one can, to a very good approximation, predict  $D_{\rm BD}/D_0$ for this system by simply substituting its $D_{\rm MD}$ data of panel Fig.~\ref{fig:StarPred}a  into eq. \ref{eq:BDMDFit} with no adjustable parameters (i.e., using $c_1$ and $c_2$ from data in Fig.~\ref{fig:TotalCollapse}, which did \textit{not} include the star-polymer fluid). As is illustrated in Fig.~S1$\dag$, the predicted values for 88\% of the state points are within 20\% of the simulation results.   

We are not aware of another  mapping approach that can make predictions of similar accuracy for both hard and ultrasoft particle fluids. One alternative strategy,\cite{deSchepper1989Long-time, Pusey1990Analogies-between, Cohen1991Note-on-transport, Lopez-Flores2011arXiv} hypothesizes that  $\rho D_{\rm MD}/[\rho D]_0= D_{\rm BD}/D_0$, where $[\rho D]_0 = \lim_{\rho \to 0} \{\rho D_{\rm MD}\}$. Although this relationship approximately holds for simple fluids with steep repulsions (the so-called hard-sphere dynamic universality class),\cite{Lopez-Flores2011arXiv} we show that it breaks down qualitatively for fluids with ultrasoft interactions.      Specifically, Fig.~S2$\dag$  illustrates that predictions based on this hypothesis for the star-polymer system are generally very inaccurate--except for a narrow region of phase space with extremely low $\rho \sigma^3$ and high $f$--where particle overlaps are avoided.    

Note that a quantitative link between $D_{\rm RMD}$ and $D_{\rm BD}/D_0$ is easy to establish for models where both can be expressed as single-valued functions of the same static quantity. For example, one can show\cite{Gnan2010Pressure-en} that, for an IPL fluid,  $D_{\rm RMD}$ and $D_{\rm BD}/D_0$ are strictly single-valued functions of excess entropy (relative to ideal gas), $\sx$. In fact, the same is approximately true for other simple liquids that are ``strongly correlating" and mimic a variety of static and dynamic properties of IPL systems.\cite{Gnan2010Pressure-en} 

Do Gaussian-core, Hertzian, and star-polymer fluids show excess-entropy scaling behaviors similar to the IPL fluids? To check this, we compute $\sx$ for these models using free-energy-based simulation methods. Specifically, we determine the density dependence of the Helmholtz free energy at high temperature using grand canonical transition-matrix Monte Carlo simulation.\cite{Errington2003Direct-calculat}  We then carry out canonical temperature-expanded ensemble simulations\cite{Lyubartsev1992New-approach-to} with a transition-matrix Monte Carlo algorithm\cite{Grzelak2010} to calculate the change in Helmholtz free energy with temperature at constant density. Together, these simulations provide  the  excess Helmholtz free energy and excess energy, and hence $\sx$. Additional details on these simulations can be found elsewhere in our earlier papers.\cite{Chopra2010On-the-use,Pond2011Implications-of} 

The excess entropy scaling behaviors of $1-D_{\rm BD}/D_0$ and $D_{\rm RMD}$ are plotted in Fig.~\ref{fig:sxD} for all model fluids and state points shown in Fig.~\ref{fig:TotalCollapse} and \ref{fig:StarPred}. The main point is that,  in stark contrast to the behavior of the IPL\ fluids, $1-D_{\rm BD}/D_0$ and $D_{\rm RMD}$ of the ultrasoft fluids are not (even approximately) single-valued functions of $\sx$. Hence, the success of the IPL-motivated mapping strategy between MD and BD diffusivities reported here cannot be explained by appealing to arguments about strongly-correlating fluids.

As a final note, we emphasize that this mapping has yet to be tested for systems with structural and dynamic properties that are strongly influenced by attractive interactions, a class of fluids that we plan to investigate in the near future.

T.M.T. acknowledges support of the
Welch Foundation (F-1696) and the National Science Foundation (CBET-1065357).
J. R. E. acknowledges financial support of the National Science Foundation (CBET-0828979).  M.J.P. acknowledges the support of the Thrust 2000 - Harry P.
Whitworth Endowed Graduate Fellowship in Engineering. The Texas
Advanced Computing Center (TACC), the University at Buffalo Center for Computational Research, and the Rensselaer Polytechnic Institute Computational Center for Nanotechnology Innovations provided computational resources for
this study.



\newpage

\begin{center}
\includegraphics[width=5in]{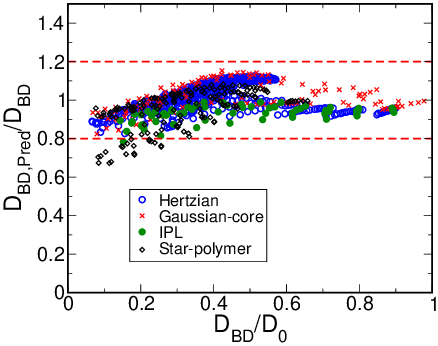}
\end{center}
{\bf Figure S1.} Ratio of long-time BD diffusivity estimated from eq. 1 in the text, $D_{\text{BD,Pred}}$, to that obtained from simulation, $D_{\text{BD}}$,  plotted as a function of  $D_{\text{BD}}/D_0$.   The dashed red lines represent a 20\% deviation of the predicted diffusivity from the value measured in simulation.
\newpage
\begin{center}
\includegraphics[width=5in]{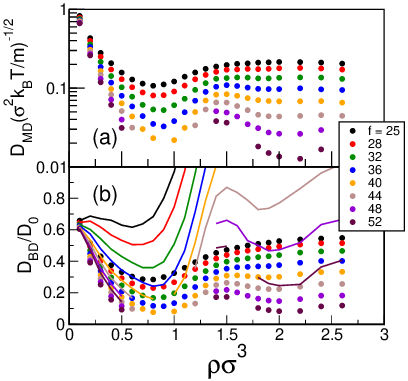}\\
\end{center}
{\bf Figure S2}. Long-time diffusivity of the star-polymer system from MD and BD simulations. Symbols are the same as in Figure 2 of the main text. Curves in panel (b) show prediction based on $D_{\text{BD}}/D_0 =\left(D_{\text{MD}}\right)/\left(\left[\rho D\right]_0 / \rho\right)$, where $\left[\rho D\right]_0 = \lim_{\rho \to 0} \rho D_{\text{MD}}$ \cite{Lopez-Flores2011arXiv}.

%

\begin{thebibliography}{40}
\expandafter\ifx\csname natexlab\endcsname\relax\def\natexlab#1{#1}\fi
\expandafter\ifx\csname bibnamefont\endcsname\relax
  \def\bibnamefont#1{#1}\fi
\expandafter\ifx\csname bibfnamefont\endcsname\relax
  \def\bibfnamefont#1{#1}\fi
\expandafter\ifx\csname citenamefont\endcsname\relax
  \def\citenamefont#1{#1}\fi
\expandafter\ifx\csname url\endcsname\relax
  \def\url#1{\texttt{#1}}\fi
\expandafter\ifx\csname urlprefix\endcsname\relax\def\urlprefix{URL }\fi
\providecommand{\bibinfo}[2]{#2}
\providecommand{\eprint}[2][]{\url{#2}}

\bibitem[{\citenamefont{Likos}(2001)}]{Likos2001effective-interactions}
\bibinfo{author}{\bibfnamefont{C.~N.} \bibnamefont{Likos}},
  \bibinfo{journal}{Phys. Rep.} \textbf{\bibinfo{volume}{348}},
  \bibinfo{pages}{267 } (\bibinfo{year}{2001}).

\bibitem[{\citenamefont{L\"{o}wen}(1994)}]{Lowen1994Melting-freezin}
\bibinfo{author}{\bibfnamefont{H.}~\bibnamefont{L\"{o}wen}},
  \bibinfo{journal}{Phys. Rep.} \textbf{\bibinfo{volume}{237}},
  \bibinfo{pages}{249} (\bibinfo{year}{1994}).

\bibitem[{\citenamefont{Ermak}(1975)}]{Ermak1975A-Computer-Sim}
\bibinfo{author}{\bibfnamefont{D.~L.} \bibnamefont{Ermak}},
  \bibinfo{journal}{J. Chem. Phys.} \textbf{\bibinfo{volume}{62}},
  \bibinfo{pages}{4189} (\bibinfo{year}{1975}).

\bibitem[{\citenamefont{Brady and Bossis}(1985)}]{Brady1985The-rheology}
\bibinfo{author}{\bibfnamefont{J.~F.} \bibnamefont{Brady}} \bibnamefont{and}
  \bibinfo{author}{\bibfnamefont{G.}~\bibnamefont{Bossis}},
  \bibinfo{journal}{J. Fluid Mech.} \textbf{\bibinfo{volume}{155}},
  \bibinfo{pages}{105} (\bibinfo{year}{1985}).

\bibitem[{\citenamefont{Hoogerbrugge and
  Koelman}(1992)}]{Hoogerbrugge1992Simulating-Micr}
\bibinfo{author}{\bibfnamefont{P.~J.} \bibnamefont{Hoogerbrugge}}
  \bibnamefont{and} \bibinfo{author}{\bibfnamefont{J.~M. V.~A.}
  \bibnamefont{Koelman}}, \bibinfo{journal}{Europhys. Lett.}
  \textbf{\bibinfo{volume}{19}}, \bibinfo{pages}{155} (\bibinfo{year}{1992}).

\bibitem[{\citenamefont{Malevanets and
  Kapral}(1999)}]{Malevanets1999Mesoscopic-mode}
\bibinfo{author}{\bibfnamefont{A.}~\bibnamefont{Malevanets}} \bibnamefont{and}
  \bibinfo{author}{\bibfnamefont{R.}~\bibnamefont{Kapral}},
  \bibinfo{journal}{J. Chem. Phys} \textbf{\bibinfo{volume}{110}},
  \bibinfo{pages}{8605} (\bibinfo{year}{1999}).

\bibitem[{\citenamefont{Heyes and
  Bra\'nka}(1994)}]{Heyes1994Molecular-and-Brownian}
\bibinfo{author}{\bibfnamefont{D.}~\bibnamefont{Heyes}} \bibnamefont{and}
  \bibinfo{author}{\bibfnamefont{A.}~\bibnamefont{Bra\'nka}},
  \bibinfo{journal}{Physics and Chemistry of Liquids}
  \textbf{\bibinfo{volume}{28}}, \bibinfo{pages}{95} (\bibinfo{year}{1994}).

\bibitem[{\citenamefont{L\"owen et~al.}(1991)\citenamefont{L\"owen, Hansen, and
  Roux}}]{Lowen1991Brownian-dynami}
\bibinfo{author}{\bibfnamefont{H.}~\bibnamefont{L\"owen}},
  \bibinfo{author}{\bibfnamefont{J.-P.} \bibnamefont{Hansen}},
  \bibnamefont{and} \bibinfo{author}{\bibfnamefont{J.-N.} \bibnamefont{Roux}},
  \bibinfo{journal}{Phys. Rev. A} \textbf{\bibinfo{volume}{44}},
  \bibinfo{pages}{1169} (\bibinfo{year}{1991}).

\bibitem[{\citenamefont{Gleim et~al.}(1998)\citenamefont{Gleim, Kob, and
  Binder}}]{Gleim1998How-Does-the-Re}
\bibinfo{author}{\bibfnamefont{T.}~\bibnamefont{Gleim}},
  \bibinfo{author}{\bibfnamefont{W.}~\bibnamefont{Kob}}, \bibnamefont{and}
  \bibinfo{author}{\bibfnamefont{K.}~\bibnamefont{Binder}},
  \bibinfo{journal}{Phys. Rev. Lett.} \textbf{\bibinfo{volume}{81}},
  \bibinfo{pages}{4404} (\bibinfo{year}{1998}).

\bibitem[{\citenamefont{Stillinger and
  Weber}(1978)}]{Stillinger1978Study-of-meltin}
\bibinfo{author}{\bibfnamefont{F.~H.} \bibnamefont{Stillinger}}
  \bibnamefont{and} \bibinfo{author}{\bibfnamefont{T.~A.} \bibnamefont{Weber}},
  \bibinfo{journal}{J. Chem. Phys.} \textbf{\bibinfo{volume}{68}},
  \bibinfo{pages}{3837} (\bibinfo{year}{1978}).

\bibitem[{\citenamefont{Foffi et~al.}(2002)\citenamefont{Foffi, Dawson,
  Buldyrev, Sciortino, Zaccarelli, and Tartaglia}}]{Foffi2002Evidence-for-an}
\bibinfo{author}{\bibfnamefont{G.}~\bibnamefont{Foffi}},
  \bibinfo{author}{\bibfnamefont{K.~A.} \bibnamefont{Dawson}},
  \bibinfo{author}{\bibfnamefont{S.~V.} \bibnamefont{Buldyrev}},
  \bibinfo{author}{\bibfnamefont{F.}~\bibnamefont{Sciortino}},
  \bibinfo{author}{\bibfnamefont{E.}~\bibnamefont{Zaccarelli}},
  \bibnamefont{and}
  \bibinfo{author}{\bibfnamefont{P.}~\bibnamefont{Tartaglia}},
  \bibinfo{journal}{Phys. Rev. E} \textbf{\bibinfo{volume}{65}},
  \bibinfo{pages}{050802(R)} (\bibinfo{year}{2002}).

\bibitem[{\citenamefont{Puertas et~al.}(2003)\citenamefont{Puertas, Fuchs, and
  Cates}}]{Puertas2003Simulation-stud}
\bibinfo{author}{\bibfnamefont{A.~M.} \bibnamefont{Puertas}},
  \bibinfo{author}{\bibfnamefont{M.}~\bibnamefont{Fuchs}}, \bibnamefont{and}
  \bibinfo{author}{\bibfnamefont{M.~E.} \bibnamefont{Cates}},
  \bibinfo{journal}{Phys. Rev. E} \textbf{\bibinfo{volume}{67}},
  \bibinfo{pages}{031406} (\bibinfo{year}{2003}).

\bibitem[{\citenamefont{Mausbach and May}(2006)}]{Mausbach2006Static-and-dyna}
\bibinfo{author}{\bibfnamefont{P.}~\bibnamefont{Mausbach}} \bibnamefont{and}
  \bibinfo{author}{\bibfnamefont{H.~O.} \bibnamefont{May}},
  \bibinfo{journal}{Fluid Phase Equilib.} \textbf{\bibinfo{volume}{249}},
  \bibinfo{pages}{17} (\bibinfo{year}{2006}).

\bibitem[{\citenamefont{Moreno and Likos}(2007)}]{Moreno2007-Cluster}
\bibinfo{author}{\bibfnamefont{A.~J.} \bibnamefont{Moreno}} \bibnamefont{and}
  \bibinfo{author}{\bibfnamefont{C.~N.} \bibnamefont{Likos}},
  \bibinfo{journal}{Phys. Rev. Lett.} \textbf{\bibinfo{volume}{99}},
  \bibinfo{pages}{107801} (\bibinfo{year}{2007}).

\bibitem[{\citenamefont{Krekelberg
  et~al.}(2009{\natexlab{a}})\citenamefont{Krekelberg, Kumar, Mittal,
  Errington, and Truskett}}]{Krekelberg2009Gaussian-dynamics}
\bibinfo{author}{\bibfnamefont{W.~P.} \bibnamefont{Krekelberg}},
  \bibinfo{author}{\bibfnamefont{T.}~\bibnamefont{Kumar}},
  \bibinfo{author}{\bibfnamefont{J.}~\bibnamefont{Mittal}},
  \bibinfo{author}{\bibfnamefont{J.~R.} \bibnamefont{Errington}},
  \bibnamefont{and} \bibinfo{author}{\bibfnamefont{T.~M.}
  \bibnamefont{Truskett}}, \bibinfo{journal}{Phys. Rev. E}
  \textbf{\bibinfo{volume}{79}}, \bibinfo{pages}{031203}
  (\bibinfo{year}{2009}{\natexlab{a}}).

\bibitem[{\citenamefont{Pamies et~al.}(2009)\citenamefont{Pamies, Cacciuto, and
  Frenkel}}]{Pamies2009Phase-diagram}
\bibinfo{author}{\bibfnamefont{J.}~\bibnamefont{Pamies}},
  \bibinfo{author}{\bibfnamefont{A.}~\bibnamefont{Cacciuto}}, \bibnamefont{and}
  \bibinfo{author}{\bibfnamefont{D.}~\bibnamefont{Frenkel}},
  \bibinfo{journal}{J. Chem. Phys.} \textbf{\bibinfo{volume}{131}},
  \bibinfo{pages}{044514} (\bibinfo{year}{2009}).

\bibitem[{\citenamefont{Pond et~al.}(2009)\citenamefont{Pond, Krekelberg, Shen,
  Errington, and Truskett}}]{Pond2009-Composition-and-conc}
\bibinfo{author}{\bibfnamefont{M.~J.} \bibnamefont{Pond}},
  \bibinfo{author}{\bibfnamefont{W.~P.} \bibnamefont{Krekelberg}},
  \bibinfo{author}{\bibfnamefont{V.~K.} \bibnamefont{Shen}},
  \bibinfo{author}{\bibfnamefont{J.~R.} \bibnamefont{Errington}},
  \bibnamefont{and} \bibinfo{author}{\bibfnamefont{T.~M.}
  \bibnamefont{Truskett}}, \bibinfo{journal}{J. Chem. Phys.}
  \textbf{\bibinfo{volume}{131}}, \bibinfo{pages}{161101}
  (\bibinfo{year}{2009}).

\bibitem[{\citenamefont{Berthier et~al.}(2010)\citenamefont{Berthier, Moreno,
  and Szamel}}]{Berthier2010arXiv-Increasing}
\bibinfo{author}{\bibfnamefont{L.}~\bibnamefont{Berthier}},
  \bibinfo{author}{\bibfnamefont{A.~J.} \bibnamefont{Moreno}},
  \bibnamefont{and} \bibinfo{author}{\bibfnamefont{G.}~\bibnamefont{Szamel}},
  \bibinfo{journal}{Phys. Rev. E} \textbf{\bibinfo{volume}{82}},
  \bibinfo{pages}{060501} (\bibinfo{year}{2010}).

\bibitem[{\citenamefont{Rosenfeld}(1977)}]{Rosenfeld1977Relation-betwee}
\bibinfo{author}{\bibfnamefont{Y.}~\bibnamefont{Rosenfeld}},
  \bibinfo{journal}{Phys. Rev. A} \textbf{\bibinfo{volume}{15}},
  \bibinfo{pages}{2545} (\bibinfo{year}{1977}).

\bibitem[{\citenamefont{Hoover}(1991)}]{Hoover1991Computational-S}
\bibinfo{author}{\bibfnamefont{W.~G.} \bibnamefont{Hoover}},
  \emph{\bibinfo{title}{Computational Statistical Mechanics}}
  (\bibinfo{publisher}{Elsevier Science Pub Co}, \bibinfo{year}{1991}), pp.
  \bibinfo{pages}{172--173}.

\bibitem[{\citenamefont{Gnan et~al.}(2009)\citenamefont{Gnan, Schr{\o}der,
  Pedersen, Bailey, and Dyre}}]{Gnan2010Pressure-en}
\bibinfo{author}{\bibfnamefont{N.}~\bibnamefont{Gnan}},
  \bibinfo{author}{\bibfnamefont{T.~B.} \bibnamefont{Schr{\o}der}},
  \bibinfo{author}{\bibfnamefont{U.~R.} \bibnamefont{Pedersen}},
  \bibinfo{author}{\bibfnamefont{N.~P.} \bibnamefont{Bailey}},
  \bibnamefont{and} \bibinfo{author}{\bibfnamefont{J.~C.} \bibnamefont{Dyre}},
  \bibinfo{journal}{J. Chem. Phys.} \textbf{\bibinfo{volume}{131}},
  \bibinfo{pages}{234504} (\bibinfo{year}{2009}).

\bibitem[{\citenamefont{Rosenfeld}(1999)}]{Rosenfeld1999A-quasi-univers}
\bibinfo{author}{\bibfnamefont{Y.}~\bibnamefont{Rosenfeld}},
  \bibinfo{journal}{J. Phys.: Condens. Matter} \textbf{\bibinfo{volume}{11}},
  \bibinfo{pages}{5415} (\bibinfo{year}{1999}).

\bibitem[{\citenamefont{Lange et~al.}(2009)\citenamefont{Lange, Caballero,
  Puertas, and Fuchs}}]{Lange2009Comparison-of-s}
\bibinfo{author}{\bibfnamefont{E.}~\bibnamefont{Lange}},
  \bibinfo{author}{\bibfnamefont{J.~B.} \bibnamefont{Caballero}},
  \bibinfo{author}{\bibfnamefont{A.~M.} \bibnamefont{Puertas}},
  \bibnamefont{and} \bibinfo{author}{\bibfnamefont{M.}~\bibnamefont{Fuchs}},
  \bibinfo{journal}{J. Chem. Phys} \textbf{\bibinfo{volume}{130}},
  \bibinfo{pages}{174903} (\bibinfo{year}{2009}).

\bibitem[{\citenamefont{Pond et~al.}(2011{\natexlab{a}})\citenamefont{Pond,
  Errington, and Truskett}}]{Pond2011Generalizing-Rosenfelds}
\bibinfo{author}{\bibfnamefont{M.~J.} \bibnamefont{Pond}},
  \bibinfo{author}{\bibfnamefont{J.~R.} \bibnamefont{Errington}},
  \bibnamefont{and} \bibinfo{author}{\bibfnamefont{T.~M.}
  \bibnamefont{Truskett}}, \bibinfo{journal}{J. Chem. Phys}
  \textbf{\bibinfo{volume}{134}}, \bibinfo{pages}{081101}
  (\bibinfo{year}{2011}{\natexlab{a}}).

\bibitem[{\citenamefont{Stillinger}(1976)}]{Stillinger1976Phase-transitio}
\bibinfo{author}{\bibfnamefont{F.~H.} \bibnamefont{Stillinger}},
  \bibinfo{journal}{J. Chem. Phys.} \textbf{\bibinfo{volume}{65}},
  \bibinfo{pages}{3968} (\bibinfo{year}{1976}).

\bibitem[{\citenamefont{Likos et~al.}(1998)\citenamefont{Likos, L\"owen,
  Watzlawek, Abbas, Jucknischke, Allgaier, and
  Richter}}]{Likos1998Star-Polymers}
\bibinfo{author}{\bibfnamefont{C.~N.} \bibnamefont{Likos}},
  \bibinfo{author}{\bibfnamefont{H.}~\bibnamefont{L\"owen}},
  \bibinfo{author}{\bibfnamefont{M.}~\bibnamefont{Watzlawek}},
  \bibinfo{author}{\bibfnamefont{B.}~\bibnamefont{Abbas}},
  \bibinfo{author}{\bibfnamefont{O.}~\bibnamefont{Jucknischke}},
  \bibinfo{author}{\bibfnamefont{J.}~\bibnamefont{Allgaier}}, \bibnamefont{and}
  \bibinfo{author}{\bibfnamefont{D.}~\bibnamefont{Richter}},
  \bibinfo{journal}{Phys. Rev. Lett.} \textbf{\bibinfo{volume}{80}},
  \bibinfo{pages}{4450} (\bibinfo{year}{1998}).

\bibitem[{\citenamefont{Foffi et~al.}(2003)\citenamefont{Foffi, Sciortino,
  Tartaglia, Zaccarelli, Verso, Reatto, Dawson, and
  Likos}}]{Foffi2003Structural-Arre}
\bibinfo{author}{\bibfnamefont{G.}~\bibnamefont{Foffi}},
  \bibinfo{author}{\bibfnamefont{F.}~\bibnamefont{Sciortino}},
  \bibinfo{author}{\bibfnamefont{P.}~\bibnamefont{Tartaglia}},
  \bibinfo{author}{\bibfnamefont{E.}~\bibnamefont{Zaccarelli}},
  \bibinfo{author}{\bibfnamefont{F.~L.} \bibnamefont{Verso}},
  \bibinfo{author}{\bibfnamefont{L.}~\bibnamefont{Reatto}},
  \bibinfo{author}{\bibfnamefont{K.~A.} \bibnamefont{Dawson}},
  \bibnamefont{and} \bibinfo{author}{\bibfnamefont{C.~N.} \bibnamefont{Likos}},
  \bibinfo{journal}{Phys. Rev. Lett.} \textbf{\bibinfo{volume}{90}},
  \bibinfo{pages}{238301} (\bibinfo{year}{2003}).

\bibitem[{\citenamefont{Allen and Tildesley}(1987)}]{Allen1987Computer-Simula}
\bibinfo{author}{\bibfnamefont{M.~P.} \bibnamefont{Allen}} \bibnamefont{and}
  \bibinfo{author}{\bibfnamefont{D.~J.} \bibnamefont{Tildesley}},
  \emph{\bibinfo{title}{Computer Simulations of Liquids}}
  (\bibinfo{publisher}{Oxford University Press, New York},
  \bibinfo{year}{1987}).

\bibitem[{\citenamefont{Wensink et~al.}(2008)\citenamefont{Wensink, L\"{o}wen,
  Rex, Likos, and van Teeffelen}}]{Wensink2008Long-time-self-}
\bibinfo{author}{\bibfnamefont{H.}~\bibnamefont{Wensink}},
  \bibinfo{author}{\bibfnamefont{H.}~\bibnamefont{L\"{o}wen}},
  \bibinfo{author}{\bibfnamefont{M.}~\bibnamefont{Rex}},
  \bibinfo{author}{\bibfnamefont{C.}~\bibnamefont{Likos}}, \bibnamefont{and}
  \bibinfo{author}{\bibfnamefont{S.}~\bibnamefont{van Teeffelen}},
  \bibinfo{journal}{Comput. Phys. Commun.} \textbf{\bibinfo{volume}{179}},
  \bibinfo{pages}{77 } (\bibinfo{year}{2008}).

\bibitem[{\citenamefont{Krekelberg
  et~al.}(2009{\natexlab{b}})\citenamefont{Krekelberg, Pond, Goel, Shen,
  Errington, and Truskett}}]{Krekelberg2009Generalized-Rosenfeld}
\bibinfo{author}{\bibfnamefont{W.}~\bibnamefont{Krekelberg}},
  \bibinfo{author}{\bibfnamefont{M.}~\bibnamefont{Pond}},
  \bibinfo{author}{\bibfnamefont{G.}~\bibnamefont{Goel}},
  \bibinfo{author}{\bibfnamefont{V.}~\bibnamefont{Shen}},
  \bibinfo{author}{\bibfnamefont{J.}~\bibnamefont{Errington}},
  \bibnamefont{and} \bibinfo{author}{\bibfnamefont{T.}~\bibnamefont{Truskett}},
  \bibinfo{journal}{Phys. Rev. E} \textbf{\bibinfo{volume}{80}},
  \bibinfo{pages}{61205} (\bibinfo{year}{2009}{\natexlab{b}}).

\bibitem[{\citenamefont{L\"owen et~al.}(1993)\citenamefont{L\"owen, Palberg,
  and Simon}}]{Lowen1993Dynamical-crite}
\bibinfo{author}{\bibfnamefont{H.}~\bibnamefont{L\"owen}},
  \bibinfo{author}{\bibfnamefont{T.}~\bibnamefont{Palberg}}, \bibnamefont{and}
  \bibinfo{author}{\bibfnamefont{R.}~\bibnamefont{Simon}},
  \bibinfo{journal}{Phys. Rev. Lett.} \textbf{\bibinfo{volume}{70}},
  \bibinfo{pages}{1557} (\bibinfo{year}{1993}).

\bibitem[{\citenamefont{de~Schepper et~al.}(1989)\citenamefont{de~Schepper,
  Cohen, Pusey, and Lekkerkerker}}]{deSchepper1989Long-time}
\bibinfo{author}{\bibfnamefont{I.~M.} \bibnamefont{de~Schepper}},
  \bibinfo{author}{\bibfnamefont{E.~G.~D.} \bibnamefont{Cohen}},
  \bibinfo{author}{\bibfnamefont{P.~N.} \bibnamefont{Pusey}}, \bibnamefont{and}
  \bibinfo{author}{\bibfnamefont{H.~N.~W.} \bibnamefont{Lekkerkerker}},
  \bibinfo{journal}{J. Phys.: Condens. Matter} \textbf{\bibinfo{volume}{1}},
  \bibinfo{pages}{6503} (\bibinfo{year}{1989}).

\bibitem[{\citenamefont{Pusey et~al.}(1990)\citenamefont{Pusey, Lekkerkerker,
  Cohen, and de~Schepper}}]{Pusey1990Analogies-between}
\bibinfo{author}{\bibfnamefont{P.}~\bibnamefont{Pusey}},
  \bibinfo{author}{\bibfnamefont{H.}~\bibnamefont{Lekkerkerker}},
  \bibinfo{author}{\bibfnamefont{E.}~\bibnamefont{Cohen}}, \bibnamefont{and}
  \bibinfo{author}{\bibfnamefont{I.}~\bibnamefont{de~Schepper}},
  \bibinfo{journal}{Physica A} \textbf{\bibinfo{volume}{164}},
  \bibinfo{pages}{12 } (\bibinfo{year}{1990}).

\bibitem[{\citenamefont{Cohen and
  de~Schepper}(1991)}]{Cohen1991Note-on-transport}
\bibinfo{author}{\bibfnamefont{E.~G.~D.} \bibnamefont{Cohen}} \bibnamefont{and}
  \bibinfo{author}{\bibfnamefont{I.~M.} \bibnamefont{de~Schepper}},
  \bibinfo{journal}{J. Stat. Phys.} \textbf{\bibinfo{volume}{63}},
  \bibinfo{pages}{241} (\bibinfo{year}{1991}).

\bibitem[{\citenamefont{Lopez-Flores et~al.}(2011)\citenamefont{Lopez-Flores,
  Mendoza-Mendez, Sanchez-Diaz, Perez-Angel, Chavez-Paez, Vizcarra-Rendon, and
  Medina-Noyola}}]{Lopez-Flores2011arXiv}
\bibinfo{author}{\bibfnamefont{L.}~\bibnamefont{Lopez-Flores}},
  \bibinfo{author}{\bibfnamefont{P.}~\bibnamefont{Mendoza-Mendez}},
  \bibinfo{author}{\bibfnamefont{L.~E.} \bibnamefont{Sanchez-Diaz}},
  \bibinfo{author}{\bibfnamefont{G.}~\bibnamefont{Perez-Angel}},
  \bibinfo{author}{\bibfnamefont{M.}~\bibnamefont{Chavez-Paez}},
  \bibinfo{author}{\bibfnamefont{A.}~\bibnamefont{Vizcarra-Rendon}},
  \bibnamefont{and}
  \bibinfo{author}{\bibfnamefont{M.}~\bibnamefont{Medina-Noyola}},
  \bibinfo{journal}{arXiv:1106.2475v1}  (\bibinfo{year}{2011}).

\bibitem[{\citenamefont{Errington}(2003)}]{Errington2003Direct-calculat}
\bibinfo{author}{\bibfnamefont{J.~R.} \bibnamefont{Errington}},
  \bibinfo{journal}{J. Chem. Phys.} \textbf{\bibinfo{volume}{118}},
  \bibinfo{pages}{9915} (\bibinfo{year}{2003}).

\bibitem[{\citenamefont{Lyubartsev et~al.}(1992)\citenamefont{Lyubartsev,
  Martsinovski, Shevkunov, and
  Vorontsov-Velyaminov}}]{Lyubartsev1992New-approach-to}
\bibinfo{author}{\bibfnamefont{A.~P.} \bibnamefont{Lyubartsev}},
  \bibinfo{author}{\bibfnamefont{A.~A.} \bibnamefont{Martsinovski}},
  \bibinfo{author}{\bibfnamefont{S.~V.} \bibnamefont{Shevkunov}},
  \bibnamefont{and} \bibinfo{author}{\bibfnamefont{P.~N.}
  \bibnamefont{Vorontsov-Velyaminov}}, \bibinfo{journal}{J. Chem. Phys.}
  \textbf{\bibinfo{volume}{96}}, \bibinfo{pages}{1776} (\bibinfo{year}{1992}).

\bibitem[{\citenamefont{Grzelak and Errington}(2010)}]{Grzelak2010}
\bibinfo{author}{\bibfnamefont{E.~M.} \bibnamefont{Grzelak}} \bibnamefont{and}
  \bibinfo{author}{\bibfnamefont{J.~R.} \bibnamefont{Errington}},
  \bibinfo{journal}{Langmuir} \textbf{\bibinfo{volume}{26}},
  \bibinfo{pages}{13297–13304} (\bibinfo{year}{2010}).

\bibitem[{\citenamefont{Chopra et~al.}(2010)\citenamefont{Chopra, Truskett, and
  Errington}}]{Chopra2010On-the-use}
\bibinfo{author}{\bibfnamefont{R.}~\bibnamefont{Chopra}},
  \bibinfo{author}{\bibfnamefont{T.~M.} \bibnamefont{Truskett}},
  \bibnamefont{and} \bibinfo{author}{\bibfnamefont{J.~R.}
  \bibnamefont{Errington}}, \bibinfo{journal}{J. Phys. Chem. B}
  \textbf{\bibinfo{volume}{114}}, \bibinfo{pages}{10558}
  (\bibinfo{year}{2010}).

\bibitem[{\citenamefont{Pond et~al.}(2011{\natexlab{b}})\citenamefont{Pond,
  Errington, and Truskett}}]{Pond2011Implications-of}
\bibinfo{author}{\bibfnamefont{M.~J.} \bibnamefont{Pond}},
  \bibinfo{author}{\bibfnamefont{J.~R.} \bibnamefont{Errington}},
  \bibnamefont{and} \bibinfo{author}{\bibfnamefont{T.~M.}
  \bibnamefont{Truskett}}, \bibinfo{journal}{arXiv:1107.4996}
  (\bibinfo{year}{2011}{\natexlab{b}}).

\end{thebibliography}

\end{document}